\documentclass[sigconf]{acmart}
\usepackage{graphicx}  
\usepackage{enumitem}  
\usepackage{xcolor}
\usepackage{listings}
\usepackage{xspace}
\usepackage{makecell}
\usepackage{graphicx}
\usepackage{multirow}
\usepackage{booktabs}
\usepackage[]{collab}

\graphicspath{{figures/}}
\newtheorem{definition}{Definition}
\newcommand\alias{\textsc{LogNLQ}\xspace}

\collabAuthor{zb}{teal}{Zhuangbin}  

\newcommand{\runinhead}[1]{\par\indent\textbf{#1.}\ }

\lstdefinestyle{sqlblock}{
  basicstyle=\ttfamily\small,
  frame=single,
  framerule=0.3pt,
  rulecolor=\color{black!25},
  backgroundcolor=\color{black!2},
  breaklines=true,
  columns=fullflexible,
  keepspaces=true,
  showstringspaces=false,
  xleftmargin=0.5em,
  xrightmargin=0.5em
}

\AtBeginDocument{%
  }

\setcopyright{acmlicensed}
\copyrightyear{2018}
\acmYear{2018}
\acmDOI{XXXXXXX.XXXXXXX}
\acmConference[Conference acronym 'XX]{Make sure to enter the correct
  conference title from your rights confirmation email}{June 03--05,
  2018}{Woodstock, NY}
\acmISBN{978-1-4503-XXXX-X/2018/06}

\begin{document}

\title{LogNLQ: Natural-Language Log Querying with Parser-Induced and Semantically Grounded Schemas}

\author{Juepeng Wang}
\affiliation{%
  \institution{Sun Yat-sen University}
  \city{Zhuhai}
  \country{China}}
\email{wangjp39@mail2.sysu.edu.cn}

\author{Jinyang Liu}
\affiliation{%
  \institution{The Chinese University of Hong Kong}
  \city{Hong Kong}
  \country{China}}
\email{jyliu@cse.cuhk.edu.hk}

\author{Zhuangbin Chen}
\authornote{Zhuangbin Chen is the corresponding author.}
\affiliation{%
  \institution{Sun Yat-sen University}
  \city{Zhuhai}
  \country{China}}
\email{chenzhb36@mail.sysu.edu.cn}
\orcid{0000-0001-5158-6716}

\author{Zibin Zheng}
\affiliation{%
  \institution{Sun Yat-sen University}
  \city{Zhuhai}
  \country{China}}
\email{zhzibin@mail.sysu.edu.cn}
\orcid{0000-0001-7872-7718}


\begin{abstract}
Logs are essential for system monitoring and failure diagnoses in modern software systems, yet querying them through natural language remains an open challenge. Existing approaches either treat logs as plain text, generate queries for schema-light backends, or assume predefined relational schemas, but none addresses a fundamental obstacle: raw logs carry no executable schema over which structured queries can be defined and run. To address these limitations, we present LogNLQ, a framework that formulates natural-language log querying as executable SQL generation over parser-induced and semantically grounded schemas. LogNLQ parses raw logs into template-partitioned relational tables, then applies dual-granularity semantic grounding to annotate both templates and parameter columns with interpretable names and descriptions. At query time, relevant schema candidates are retrieved via semantic search, and a large language model (LLM) generates executable SQL constrained to the retrieved context. To support rigorous evaluation, we introduce LogNLQ-Bench, an execution-verified benchmark of 8,895 queries over four real-world log datasets. Experimental results demonstrate that LogNLQ consistently outperforms all representative baselines by wide margins, with especially pronounced gains on analytically complex scenario queries.
\end{abstract}



\maketitle

\section{Introduction}
\label{sec:intro}

Logs are a foundational source of runtime observability in modern software systems~\cite{DBLP:journals/cacm/OlinerGX12, DBLP:journals/csur/HeHCYSL21}, routinely produced at scales of millions to billions of lines per day. Engineers rely on them to diagnose failures, inspect anomalies~\cite{chen2021experience}, audit system behavior, and understand operational trends~\cite{he2022empirical, du2017deeplog, he2016experience}. Although modern observability platforms provide powerful search and analytics capabilities, these are typically exposed through backend-specific query languages, operators, and ad-hoc regular expressions, requiring users to translate naturally expressed diagnostic intent into low-level query formulations.

Natural-language querying (NLQ) is therefore a promising direction for closing this gap. Existing approaches primarily fall into three categories (Section~\ref{sec:background}), each capturing part of the solution but missing a critical piece. \emph{Text-based Log QA} methods~\cite{huang2023logqa, qi2024logsay} retrieve relevant log snippets and extract local answer spans, but they operate at the granularity of individual lines and cannot faithfully execute corpus-level operations such as counting, grouping, or filtering by parameter role. \emph{Log-DSL Generation} methods~\cite{seshagiri2024chatting, tang2024nl2kql} translate natural language into backend-native query languages such as LogQL~\cite{logql_manual} or KQL~\cite{kql_docs}, which are in principle executable; in practice, however, raw logs expose no stable schema, so models are forced to guess field names and frequently produce hallucinated or empty results~\cite{DBLP:journals/csur/JiLFYSXIBMF23}. \emph{Text-to-SQL for Logs}~\cite{pourreza2023din, yu2018spider} is the most naturally aligned with analytical querying, as SQL natively supports filtering, aggregation, and grouping. However, it presumes the existence of stable, human-authored schemas with semantically meaningful tables and columns, an assumption that raw logs fundamentally violate. The core difficulty shared by all three is therefore not translating natural language into queries, but first establishing the executable schema that raw logs do not provide.

This analysis reveals a fundamental prerequisite that all three paradigms overlook: before any query can be generated, an \textit{executable} schema must first be induced from raw logs. Although logs lack predefined schemas, they are not entirely unstructured. Log messages are produced by logging statements in source code~\cite{DBLP:conf/icse/YuanPZ12, DBLP:conf/kbse/HeCHL18, DBLP:conf/kbse/LiLPJCHZCLSL25}, generating recurring textual templates whose dynamic segments carry typed runtime values (e.g., IP addresses, error codes, resource identifiers). This regularity makes it possible to parse raw logs into structured event templates and explicit parameter columns, yielding a relational substrate over which SQL queries can be precisely defined and executed. Structural induction alone, however, is not sufficient. The resulting parameter columns are positional placeholders that the execution engine can process but that an LLM cannot reliably map to user-facing concepts such as ``source IP.'' Semantic grounding is therefore needed to annotate induced schema elements with meaningful names and descriptions, completing the bridge between raw log structure and natural-language intent.

Based on this analysis, we present \textbf{LogNLQ}, a framework that formulates natural-language log querying as executable SQL generation over parser-induced and semantically grounded schemas. In the offline phase, LogNLQ parses raw logs into template-partitioned relational tables and applies dual-granularity semantic grounding to annotate both event templates and individual parameter columns with interpretable labels and descriptions. In the online phase, given a user query, LogNLQ retrieves the most relevant schema candidates through semantic search, constructs a grounded prompt, and directs an LLM to generate executable SQL constrained to the retrieved schema context. To support rigorous evaluation, we also introduce \textbf{LogNLQ-Bench}, an execution-verified benchmark of 8,895 natural-language log queries over four real-world datasets, covering both retrieval and aggregation tasks under basic and scenario-driven settings. Experiments on LogNLQ-Bench show that LogNLQ consistently outperforms all representative baselines by wide margins, achieving retrieval F1 scores above 0.89 on basic queries where the strongest baselines remain below 0.25, with especially pronounced gains on analytically complex scenario queries.

The contributions of this paper are summarized as follows:

\begin{itemize}[noitemsep,leftmargin=5.5mm]
    \item \textbf{Problem reformulation.} We formulate natural-language log querying as executable query generation over \emph{induced} rather than pre-authored schemas, highlighting the schema-missing assumption gap that prevents conventional Text-to-SQL and text-based QA methods from directly handling raw logs.

    \item \textbf{Schema-induction-based framework.} We present LogNLQ, a framework that couples parsing-based structural induction with semantic grounding, schema retrieval, and constrained SQL generation to support executable natural-language querying over logs.

    \item \textbf{LogNLQ-Bench.} We introduce an execution-verified benchmark of 8,895 natural-language log queries over four real-world datasets, covering both retrieval and aggregation tasks under basic and scenario-driven settings.

    \item \textbf{Empirical evaluation.} We show that parser-induced structuring is an absolute prerequisite for executable log analytics, and that semantic grounding further yields substantial gains over representative baselines, with especially clear improvements on scenario-driven and analytically complex queries.
\end{itemize}

The remainder of this paper is organized as follows. Section~\ref{sec:background} presents background and motivating examples. Section~\ref{sec:methodology} describes the design of LogNLQ. Section~\ref{sec:experiments} introduces the benchmark, baselines, and evaluation setup. Section~\ref{sec:evaluation} reports the experimental results. Section~\ref{sec:related_work} discusses related work, and Section~\ref{sec:conclusion} concludes the paper.

\section{Background and Motivation}
\label{sec:background}

\subsection{Logs and Log Querying}

Logs are timestamped records emitted continuously by running software, capturing events such as service requests, errors, state transitions, and resource usage. They serve as a primary observability primitive across diverse operational scenarios: engineers inspect logs to diagnose failures and anomalies, security teams mine them for audit trails and intrusion detection, and SRE teams rely on them for performance analysis and capacity planning.

As log volumes have grown to billions of lines per day, the tooling for log querying has evolved accordingly. Early practice depended on command-line tools such as \texttt{grep} for manual inspection of local files. Modern systems ingest logs into centralized platforms such as Elasticsearch~\cite{elasticsearch} and Grafana Loki~\cite{grafana_loki}, which support efficient search, filtering, and aggregation at scale. Yet this evolution does not close the gap between user intent and system interface. These platforms improve storage and execution, but they still expose backend-native query languages such as LogQL and Elasticsearch Query DSL, both of which require mastery of platform-specific syntax and weakly exposed schema fields. The burden of translating diagnostic intent into executable queries remains on the user, motivating the search for natural-language log querying interfaces.

\subsection{Motivating Example}

To concretize the challenges outlined above, we trace a representative diagnostic query through three existing paradigms. Consider the request: \textit{``Count failed SSH logins from IP 10.0.0.1.''} Although this query appears straightforward, answering it correctly requires the system to jointly satisfy three conditions: identifying the correct event type (failed authentication), targeting the correct parameter role (source IP rather than some other string position), and executing a corpus-wide aggregation (\texttt{COUNT}). These conditions reveal that natural-language log querying is not merely a retrieval problem over text, but an execution problem over latent structure that raw logs do not explicitly expose.


\begin{figure}[htbp]
  \centering
  \includegraphics[width=\columnwidth]{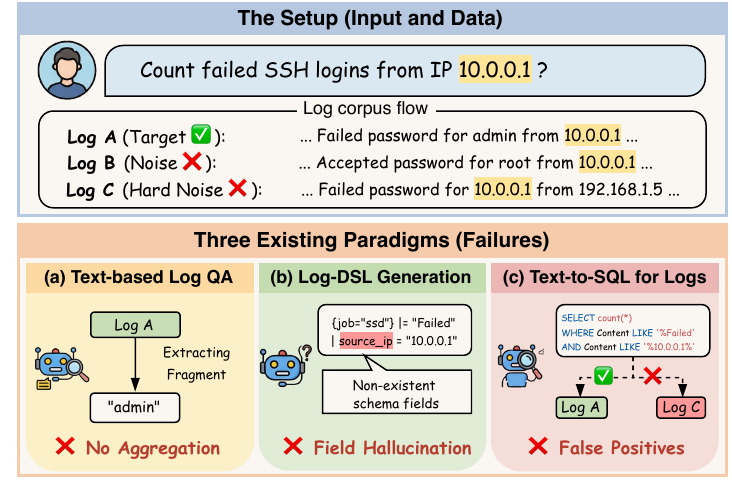}
  \caption{Motivating Example of NL Log Querying.}
  \Description{Failure modes of three NL log querying paradigms on a representative query.}
  \label{fig:motivation}
\end{figure}

As illustrated in Figure~\ref{fig:motivation}, the raw log corpus contains not only simple noise (e.g., Log B, a successful login), but also ``hard noise'' such as Log C, where the target IP (\texttt{10.0.0.1}) appears in a structurally misleading position as a username. This example exposes a common failure mode across existing paradigms. As shown in Figure~\ref{fig:motivation}(a), \emph{Text-based Log QA} methods may recover local entities from individual log lines, but they cannot faithfully execute corpus-level analytical operations such as \texttt{COUNT}. Figure~\ref{fig:motivation}(b) shows that \emph{Text-to-SQL for Logs} over monolithic raw-log storage can generate executable SQL, yet typically relies on coarse substring predicates (e.g., \texttt{LIKE}) that ignore parameter roles and therefore wrongly count Log C as a match. Figure~\ref{fig:motivation}(c) illustrates the opposite failure mode for \emph{Log-DSL Generation}: because raw logs do not expose a stable relational schema, the model may hallucinate plausible but nonexistent fields such as \texttt{source\_ip}, causing failed or empty executions. Across all three cases, the underlying problem is the same: they attempt query resolution without first establishing an executable and semantically interpretable schema over logs.

This example reveals not just what goes wrong, but \emph{why}: all three paradigms attempt query resolution against a representation that lacks the necessary structure. The root cause is that raw logs are semi-structured streams with no stable, human-authored schema, yet precise parameter-level filtering and corpus-scale aggregation both depend on exactly such structure. This points to a clear design imperative. We need, first, \emph{structural induction}: raw logs must be parsed into typed event templates and parameter columns, so that execution targets specific parameter roles rather than fragile substring patterns. We need, second, \emph{semantic grounding}: the induced schema elements (positional columns such as \texttt{p\_1} and \texttt{p\_2}) must be annotated with meaningful names and descriptions, so that an LLM can map user-facing concepts to the correct physical fields. Without the first, corpus-level aggregation collapses entirely. Without the second, the LLM cannot navigate the induced structure reliably. LogNLQ is built around this precise combination: induce a relational schema from raw logs, ground it semantically, and generate constrained SQL over the grounded schema.

\section{Methodology}
\label{sec:methodology}

\subsection{Overview}

In this section, we present the design of LogNLQ. Given a corpus of raw logs $L$ and a natural-language query $Q$, the goal is to generate an executable SQL query $S$ whose execution returns results consistent with the user's intent. Unlike conventional Text-to-SQL settings, where a semantically meaningful database schema is assumed to exist \emph{a priori}, LogNLQ must first induce such a schema from raw logs before query generation can proceed. LogNLQ therefore couples structural induction, semantic grounding, schema retrieval, and constrained SQL generation into a single querying workflow.


\begin{figure*}[t]
  \centering
  \includegraphics[width=0.62\textwidth]{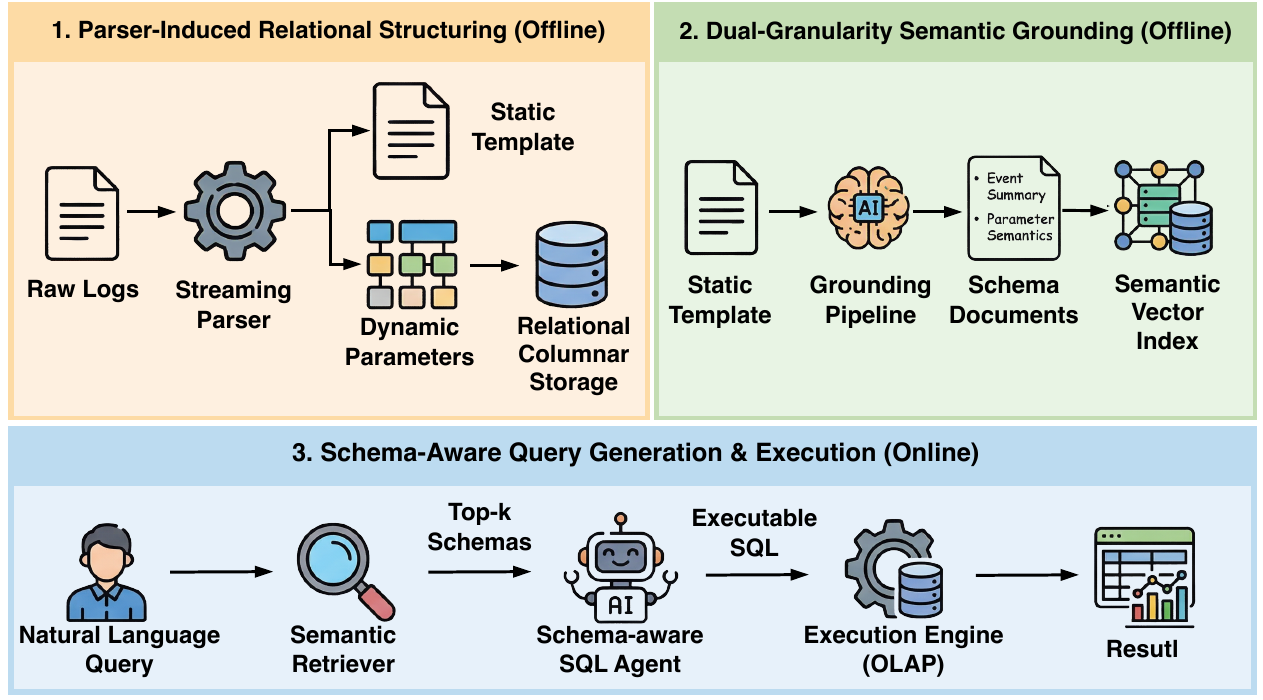}
  \caption{Overview of the LogNLQ Pipeline.}
  \Description{LogNLQ pipeline: offline schema induction and indexing; online query execution.}
  \label{fig:overview}
\end{figure*}

As illustrated in Figure~\ref{fig:overview}, LogNLQ comprises an offline preparation phase and an online querying phase that together form a unified schema-synthesis workflow. In the offline phase, raw logs are processed by the schema induction layer into event templates and dynamic parameters, materialized into relational storage, semantically grounded, and indexed as schema documents. In the online phase, a user query retrieves the most relevant schema candidates, an LLM generates SQL under the retrieved schema context, and the query is executed over the structured log store.


The interface between the two phases is a set of \emph{schema documents}, which serve as the final induced schema abstraction exposed to retrieval and SQL generation. Each schema document encapsulates not only the parser-induced physical structure of one event template, but also the semantic information required to make that structure understandable to the LLM. We define this central data structure formally as follows.

\begin{definition}[Schema Document]
A schema document $\mathcal{D}_k$ for template $T_k$ is a tuple $\langle e_k,\, T_k,\, s_k,\, \{(c_i, n_i, d_i, \sigma_i)\}_{i=1}^{m} \rangle$, where $e_k$ is the event identifier, $T_k$ is the template string, $s_k$ is a natural-language summary of the event semantics, and each $(c_i, n_i, d_i, \sigma_i)$ describes one parameter column by its physical name $c_i$ (e.g., \texttt{p\_1}), canonical semantic name $n_i$ (e.g., \textit{Source IP}), natural-language description $d_i$, and statistical profile $\sigma_i$ (comprising value cardinality, range, and top-$K$ most frequent values).
\end{definition}

\begin{figure}[htbp]
  \centering
  \includegraphics[width=0.94\columnwidth]{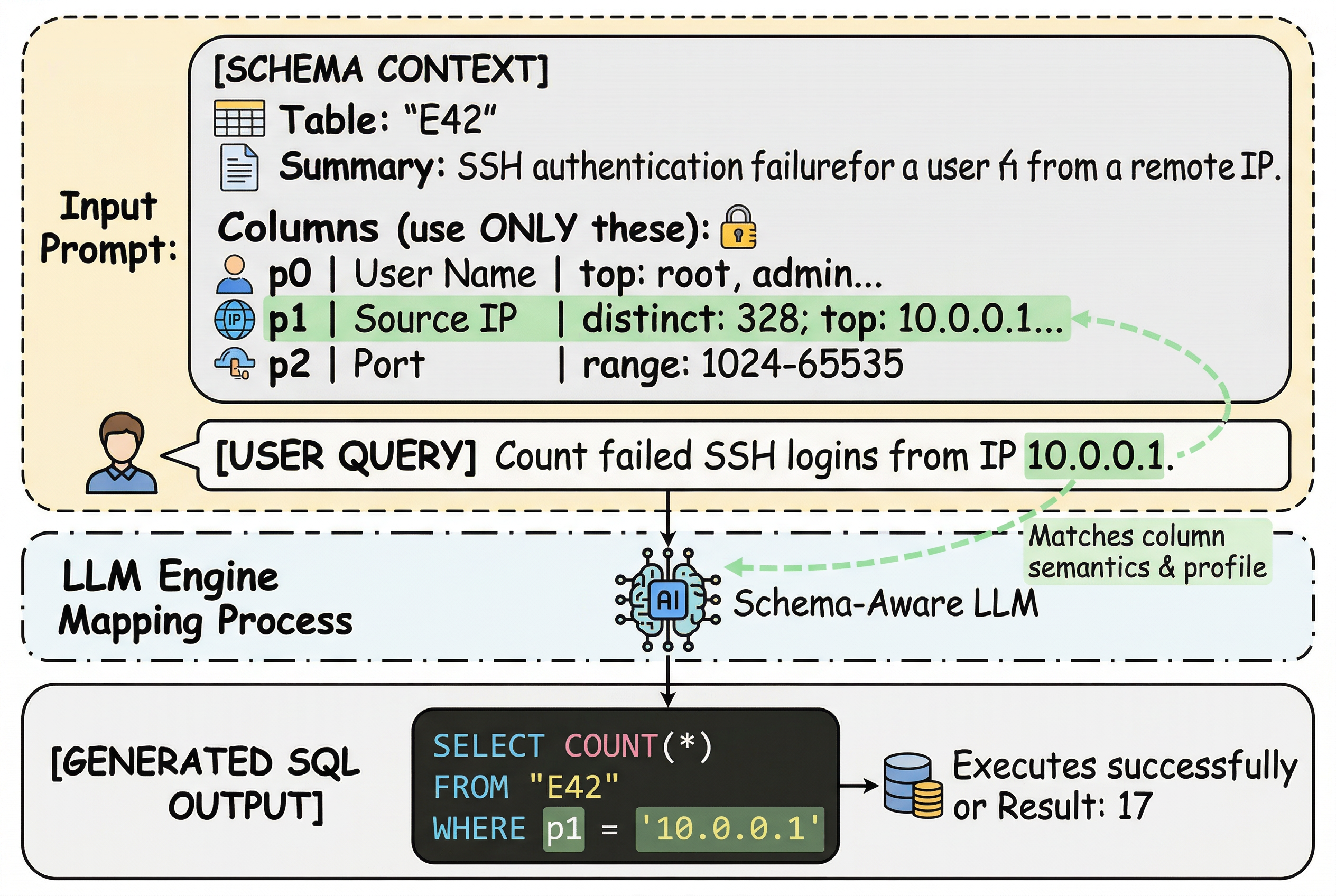}
  \caption{End-to-End Walkthrough of LogNLQ.}
  \Description{A log query traced through parsing, grounding, retrieval, and SQL generation.}
  \label{fig:running_example}
\end{figure}

Schema documents serve as both the retrieval units for online schema selection and the grounded context for SQL generation. Figure~\ref{fig:running_example} traces a concrete example through the full pipeline, which we use as a running example throughout this section.

\subsection{Execution-Oriented Schema Induction}

Schema induction establishes the \emph{physical foundation} of the LogNLQ pipeline by transforming raw logs into an executable relational representation. The key observation is that each log message can be decomposed into two parts: a recurring event template $T_k$ (the static portion shared across similar messages) and an ordered sequence of dynamic parameters $P_i = [p_{i,1}, p_{i,2}, \dots, p_{i,n}]$ (the runtime-specific values). This decomposition maps naturally to a relational model in which each distinct template defines a logical table and each parameter position defines a column within that table. LogNLQ exploits this correspondence by parsing each log line $l_i$ into a tuple $(T_k, P_i)$ and materializing the results into a template-partitioned columnar store, producing the physical substrate over which downstream SQL generation and execution operate.

\runinhead{Streaming dispatch and schema management}
Because logs arrive as a continuous stream rather than a static corpus, the schema inducer must discover and manage templates incrementally. Each incoming log line is routed through one of three paths. If the line matches an existing template $T_k$, its extracted parameters are batched and appended to the corresponding partition. If the schema inducer identifies a previously unseen message pattern, a new template is registered, a dedicated partition is created, and the semantic grounding pipeline (Section~\ref{sec:grounding}) is triggered asynchronously for the new template. If a newly stabilized template is found to carry additional parameter positions not present in earlier records, the partition schema is extended with the new columns, and previously stored records are retroactively subjected to positional parameter extraction under the updated schema, with the extracted values populated in the new columns.

\runinhead{Template-partitioned columnar storage}
To enable column-level SQL operations over the induced structure, LogNLQ materializes each template $T_k$ as an independent partition in a column-oriented store, identified by its event identifier $e_k$. The partition schema contains metadata columns (\texttt{LineId}, \texttt{Timestamp}) together with positional parameter columns \texttt{p\_1}, \texttt{p\_2}, \dots\ corresponding to the extracted positions. Each original log line is fully reconstructable by substituting the stored parameter values back into the template string, so raw log content need not be stored separately. This template-partitioned layout has two practical advantages. First, schema evolution is partition-local: modifications triggered by template refinement affect only the relevant partition, leaving all others unchanged. Second, query execution can prune irrelevant partitions entirely and access only the columns required by the predicate, enabling efficient analytical operations over large log corpora. More importantly, this materialization step turns parser output from a descriptive artifact into an executable database substrate. The induced templates are no longer merely mined patterns; they become the actual relational units over which downstream querying operates.

\runinhead{Statistical profiling}
While the induced schema provides the structural context that an LLM needs to identify correct tables and columns, generating precise filter predicates also requires knowledge of the actual values that appear in each column. For each parameter column $c_i$ of template $T_k$, LogNLQ therefore derives a lightweight statistical profile $\sigma_i$ comprising cardinality (number of distinct values), value range, and the top-$K$ most frequent values. Rather than being updated incrementally on every record write, a background process periodically recomputes profiles from each partition. Because parameter value distributions in structured logs stabilize quickly as partitions grow, profiles derived at query time reliably capture the operational data characteristics. They are supplied alongside the schema document during SQL generation to guide precise filter construction.

\subsection{Dual-Granularity Semantic Grounding}
\label{sec:grounding}

The relational structure induced by parsing is executable, but it is not yet a usable schema for LLM-based query generation. Positional columns such as \texttt{p\_1} and \texttt{p\_2} expose structural boundaries without revealing semantic roles, making the induced database legible to the execution engine but largely opaque to the model. LogNLQ therefore performs semantic grounding as the \emph{logical induction layer} that complements parser-induced structure. The goal is not merely to enrich the schema with additional descriptions, but to complete the induced schema so that it becomes interpretable enough for the LLM to align user intent with physical fields.

\runinhead{Template-level grounding}
Users express diagnostic intent in terms of event semantics (e.g., ``failed login,'' ``disk error'') rather than raw template strings. To bridge this gap, LogNLQ generates a short natural-language summary $s_k$ for each parsed template $T_k$, capturing the operational meaning of the event. The input includes the template string together with representative log instances sampled from the materialized partition. The resulting summary enables schema retrieval to connect high-level user queries with the corresponding induced table.

\runinhead{Parameter-level grounding}
Similarly, users refer to specific data fields by their semantic roles (e.g., ``source IP,'' ``user name'') rather than by positional indices. To make individual columns interpretable, LogNLQ infers a canonical semantic name $n_i$ and description $d_i$ for each parameter column $c_i$ of template $T_k$ by presenting an LLM with the template string and a sample of representative values. The LLM produces a space-separated canonical name following a consistent capitalization convention (e.g., \textit{Source IP}, \textit{User Name}, \textit{Port Number}) together with a one-sentence description.

A key challenge is ensuring cross-template consistency: semantically equivalent parameters may appear under different positional indices across templates, with a source IP address appearing as \texttt{p\_1} in one template and \texttt{p\_2} in another. If these receive different labels, retrieval may miss relevant templates and cross-template SQL generation (e.g., UNION operations) cannot reliably align columns across partitions. To address this, LogNLQ maintains a \emph{global parameter vocabulary} $\mathcal{G}$ that accumulates canonical name-description pairs across all processed templates. When grounding a new column, rather than injecting the full $\mathcal{G}$ into the prompt, LogNLQ retrieves the most semantically similar entries from $\mathcal{G}$ via lightweight vector search and provides only this focused subset as context. The LLM is instructed to reuse an existing entry if one is semantically equivalent, or to introduce a new entry otherwise. This vocabulary-anchored grounding ensures that parameters sharing the same operational role receive consistent labels regardless of their positional index, enabling retrieval and SQL generation to reason over semantic roles rather than positional proxies.

\runinhead{Schema document construction and indexing}
Because user queries may target either event-type semantics (``failed SSH login'') or specific parameter roles (``source IP''), the retrieval index must capture both levels in a single searchable representation. For each template $T_k$, LogNLQ assembles the physical and semantic information into the schema document $\mathcal{D}_k$ defined above. The template-level summary $s_k$ and the parameter annotations $\{(c_i, n_i, d_i)\}$ are concatenated into a single composite text and encoded into a dense vector. Schema documents are indexed with FAISS, forming the retrieval layer for the online phase. The statistical profiles $\sigma_i$ are not part of the indexed vector representation; they are computed on demand at query time and attached to the retrieved schema documents to enrich the SQL generation prompt.

\subsection{Schema-Aware Query Generation and Execution}

At query time, LogNLQ translates a natural-language request into an executable SQL query through a three-step online workflow: schema retrieval, constrained SQL generation, and execution. Because the offline phase has already induced and grounded the schema, the online phase operates over a compact set of semantically annotated schema candidates rather than over raw log text.

\runinhead{Semantic schema retrieval}
The full set of induced templates may contain hundreds of event types, most of which are irrelevant to any given query. To focus the generation context on the most pertinent schemas, LogNLQ embeds the user query $Q$ and performs nearest-neighbor search over the schema-document index, retrieving the Top-$K$ candidates whose event semantics and parameter annotations are most relevant to $Q$. Unlike snippet retrieval over raw logs, this process retrieves candidate executable schemas that define the feasible space for downstream SQL generation.

\runinhead{Schema-constrained SQL generation}
Unconstrained generation over log data risks producing references to field names that do not exist in the induced schema. To prevent this, LogNLQ constructs a prompt containing all Top-$K$ retrieved schema documents and explicitly constrains the LLM to reference only the columns enumerated in these documents. The prompt enumerates each candidate's table name, template string, summary, available columns with their canonical names and descriptions, and representative statistics. In most cases the generated SQL targets a single template partition. When the query semantics require it, the LLM is explicitly prompted to generate SQL spanning multiple candidate partitions via UNION or JOIN operations. This is made tractable by the bounded candidate set (Top-$K$ schemas rather than an unbounded table catalog) and aided by the global parameter vocabulary, which aligns semantically equivalent fields across templates. Figure~\ref{fig:running_example} illustrates how this grounded schema context guides the model to map user intent to exact physical columns during SQL generation.

\runinhead{Execution}
The generated SQL is executed over the template-partitioned store. Because the storage layout is template-aware and column-oriented, the execution engine prunes irrelevant partitions and accesses only the columns required by the query. The final result is returned to the user.

\section{Experimental Setup}
\label{sec:experiments}

We evaluate LogNLQ along three dimensions: end-to-end effectiveness, system overhead, and component contribution. This section introduces the benchmark, baselines, evaluation metrics, and implementation details.

\subsection{Benchmark Construction}

No existing benchmark directly targets natural language log querying with execution-verified analytical grounding, so we construct LogNLQ-Bench. The benchmark covers common operational querying needs while remaining independent of the LogNLQ pipeline.

Following prior empirical taxonomies of industrial log analysis, we organize the benchmark along two levels of difficulty:

\begin{itemize}[noitemsep,leftmargin=5.5mm]
    \item \textbf{Basic Queries:} These queries cover common ad hoc requests such as straightforward filtering, counting, and simple aggregation.
    \item \textbf{Scenario Queries:} These queries model more involved diagnostic workflows, including alert inspection, triage, mitigation, root cause analysis, auditing, and reporting. Many scenario queries inherently span multiple event types and thus require cross-template SQL generation.
\end{itemize}

Across both difficulty levels, we consider two task categories that define the evaluation scope of LogNLQ-Bench:

\begin{itemize}[noitemsep,leftmargin=5.5mm]
    \item \textbf{Log Retrieval:} the system returns the set of log records satisfying a natural language condition.
    \item \textbf{Metric Aggregation:} the system returns a derived analytical result, such as a count, grouped summary, or other aggregate statistic computed from the log corpus.
\end{itemize}

LogNLQ-Bench is constructed through a scenario-driven, execution-verified pipeline. First, we sample raw-log windows from each dataset as the basis for question generation. Guided by the SRE operational workflows outlined in~\cite{Beyer:2016:SRE}, the LLM proposes a plausible diagnostic question based on a sampled log window and a specific scenario type, while simultaneously synthesizing an executable analysis program to derive the ground-truth answer directly from the raw logs. We then retain only examples whose generated analysis programs execute successfully and yield valid non-empty outputs. Exact and near-duplicate questions are removed, and ambiguous phrasing is refined when necessary. Finally, we manually inspect a stratified sample of 200 benchmark instances (approximately 2\% of the total) to assess linguistic naturalness, answer plausibility, and consistency with operational workflows.

We apply this pipeline to four widely used datasets from LogHub-2.0~\cite{jiang2024large}: OpenSSH, HDFS, Spark, and BGL. They span different systems and log characteristics, enabling evaluation across diverse operational domains. The final benchmark contains 8,895 queries in total. Table~\ref{tab:dataset_properties} summarizes dataset scale and benchmark composition. The 1,139 scenario queries are distributed nearly uniformly across the six types (Alert, Triage, Mitigation, RCA, Audit, and Report), with each type accounting for roughly one-sixth of the total.

\begin{table*}[htbp]
  \caption{Dataset Properties and LogNLQ-Bench Statistics}
  \label{tab:dataset_properties}
  \small
  \begin{tabular}{l c c c c c c c}
    \toprule
    & & \multicolumn{2}{c}{\textbf{Basic Queries}} & \multicolumn{2}{c}{\textbf{Scenario Queries}} \\
    \cmidrule(lr){3-4} \cmidrule(lr){5-6}
    \textbf{Dataset} & \textbf{Raw Logs} & \textbf{Retrieval} & \textbf{Aggregation} & \textbf{Retrieval} & \textbf{Aggregation} & \textbf{Total} \\
    \midrule
    \textbf{OpenSSH} & 638,947    & 116   & 158   & 133 & 141 & 548 \\
    \textbf{HDFS}    & 11,167,740 & 314   & 279   & 142 & 146 & 881 \\
    \textbf{Spark}   & 16,075,117 & 1,315 & 1,662 & 143 & 143 & 3,263 \\
    \textbf{BGL}     & 4,631,261  & 1,911 & 2,001 & 146 & 145 & 4,203 \\
    \midrule
    \textbf{Total}   & \textbf{32,513,065} & \textbf{3,656} & \textbf{4,100} & \textbf{564} & \textbf{575} & \textbf{8,895} \\
    \bottomrule
  \end{tabular}
\end{table*}

\subsection{Baseline Methods}

We compare LogNLQ against several representative systems. Our goal is to evaluate end-to-end utility under the native storage and querying assumptions of these methods rather than retrofitting them with our parser-induced representations.

\begin{itemize}[noitemsep,leftmargin=5.5mm]
    \item \textbf{LogQA}: LogQA fine-tunes a BERT model with log question-answer pairs to locate specific answer spans within matching log entries. We reproduced this method using LogNLQ-Bench pairs for the Log Retrieval task exclusively, extracting results matching the ground-truth answer length from its ranked similarity list because it lacks analytical aggregation capabilities.
    
    \item \textbf{LogSay}: LogSay is a retriever-reader system that incorporates a numerical reasoner to handle logical operations over localized text snippets. We deployed it to process our benchmark queries to evaluate its capacity for extracting or computing answers without relational aggregation.
    
    \item \textbf{Elasticsearch}: Elasticsearch is a distributed search engine utilizing inverted indices for efficient keyword-based content retrieval. We constructed an LLM-based query generator to translate user intents into native Elasticsearch syntax and executed these queries over an inverted-indexed raw log layout.
    
    \item \textbf{Loki}: Grafana Loki is a log aggregation system that evaluates LogQL queries over label-indexed log streams rather than relying on full-text indexing. We prompted a large language model to synthesize LogQL queries from natural language and executed them against a Loki backend to assess performance without explicit parameter columns.
    
    \item \textbf{LogQLLM}: LogQLLM leverages large language models to translate natural language questions directly into executable LogQL statements. We evaluated its end-to-end translation efficacy on our datasets to test its reliance on content matching and weak field inference over semi-structured logs.
    
    \item \textbf{DIN-SQL}: DIN-SQL serves as a strong baseline for Text-to-SQL tasks by decomposing query generation into structured sub-tasks like schema linking and self-correction through in-context learning. We adapted this approach to operate over a monolithic text storage layout to determine if step-by-step generation can compensate for the absence of parser-induced structures.
\end{itemize}

\subsection{Evaluation Metrics}

We evaluate generated queries using task-specific metrics for retrieval and aggregation.

\begin{itemize}[noitemsep,leftmargin=5.5mm]
    \item \textbf{F1 Score:} For retrieval queries, we compare the set of returned log identifiers (\texttt{LineId}) against the ground-truth set and report Precision, Recall, and F1-score:
    \begin{equation}
        F1 = \frac{2 \times Precision \times Recall}{Precision + Recall}.
    \end{equation}

    \item \textbf{Result-Oriented Match:} For aggregation queries, the output is a derived analytical result rather than a set of raw log instances, and multiple executable queries may legitimately satisfy the same intent. We therefore adopt a Result-Oriented Match metric (RO-Match), which evaluates the executed result rather than the surface form of the generated query.
\end{itemize}

A generated result is counted as a match if it is either numerically or structurally identical to the ground-truth result after canonicalization, or, for set-valued or grouped analytical outputs, a strict subset of the ground-truth result that contains no incorrect extra entries. Subset matching is never credited for scalar outputs such as \texttt{COUNT} or \texttt{AVG}, for which exact value equality is required. Matching is determined by a rule-based comparator over structured execution outputs, including exact value equality, canonicalized table comparison, and containment checks where applicable.

\subsection{Implementation Details}

The offline pipeline is implemented using a custom schema induction component whose streaming template extraction builds on the cache-tree dispatch mechanism of LILAC~\cite{lilac}, extended to support execution-oriented relational structuring. Dense vector representations for schema documents are generated using the \texttt{doubao-embedding} model and indexed with FAISS (CPU version). Online query execution is performed by DuckDB over Snappy-compressed Parquet files. For a fair comparison, all generative baselines and LogNLQ use the same foundation model (\texttt{doubao-seed-} \texttt{2.0-pro} via API) with generation temperature set to 0.0. All experiments are conducted on a Linux server with an Intel Xeon CPU and 32GB of RAM.

These components constitute a representative prototype implementation rather than a prescribed deployment stack. The core design of LogNLQ (parser-induced relational structuring and semantic grounding) is independent of any particular execution engine and can in principle be integrated with other local or distributed analytical backends. Unless otherwise noted, all main experiments use \texttt{doubao-seed-2.0-pro} as the shared backbone for all generative methods to ensure fair, controlled comparison. The effect of alternative backbones on LogNLQ is studied separately in RQ3.

\section{Evaluation Results}
\label{sec:evaluation}

In this section, we present the results of our evaluation, which aim to answer the following three research questions:

\begin{itemize}[noitemsep,leftmargin=5.5mm]
    \item \textbf{RQ1 (Effectiveness):} How accurately does LogNLQ execute retrieval and aggregation queries compared with representative baselines?
    \item \textbf{RQ2 (Efficiency):} What overhead does LogNLQ introduce in storage, offline preprocessing, and online latency?
    \item \textbf{RQ3 (Ablation and Sensitivity):} How much do parser-induced structuring and semantic grounding each contribute, how sensitive is the system to the number of retrieved schema candidates, and how does the choice of LLM backbone affect SQL generation accuracy?
\end{itemize}

\subsection{RQ1: Effectiveness}

Table~\ref{tab:combined_performance_merged} reports end-to-end performance on Log Retrieval (F1) and Metric Aggregation (RO-Match). LogNLQ achieves the best performance in every setting, with especially wide margins on scenario queries and aggregation tasks.

\begin{table*}[t]
  \centering
  \caption{Performance Comparison on Log Retrieval (F1) and Metric Aggregation (RO-Match)}
  \label{tab:combined_performance_merged}
  \small
  \setlength{\tabcolsep}{4pt}
  \begin{tabular}{l l c c c c c c c c}
    \toprule
    \multirow{2}{*}{\textbf{Method}} & \multirow{2}{*}{\textbf{Metric}} & \multicolumn{2}{c}{\textbf{OpenSSH}} & \multicolumn{2}{c}{\textbf{HDFS}} & \multicolumn{2}{c}{\textbf{Spark}} & \multicolumn{2}{c}{\textbf{BGL}} \\
    \cmidrule(lr){3-4} \cmidrule(lr){5-6} \cmidrule(lr){7-8} \cmidrule(lr){9-10}
    & & \textbf{Basic} & \textbf{Scenario} & \textbf{Basic} & \textbf{Scenario} & \textbf{Basic} & \textbf{Scenario} & \textbf{Basic} & \textbf{Scenario} \\
    \midrule

    \multirow{2}{*}{LogQA}
    & F1 & 0.092 & 0.161 & 0.154 & 0.132 & 0.081 & 0.045 & 0.076 & 0.051 \\
    & RO & - & - & - & - & - & - & - & - \\
    \cmidrule(lr){1-10}

    \multirow{2}{*}{LogSay}
    & F1 & 0.125 & 0.084 & 0.182 & 0.045 & 0.114 & 0.072 & 0.203 & 0.068 \\
    & RO & 0.031 & 0.006 & 0.018 & 0.004 & 0.025 & 0.008 & 0.041 & 0.009 \\
    \cmidrule(lr){1-10}

    \multirow{2}{*}{LogQLLM}
    & F1 & 0.148 & 0.104 & 0.115 & 0.072 & 0.142 & 0.095 & 0.131 & 0.092 \\
    & RO & 0.128 & 0.085 & 0.112 & 0.062 & 0.118 & 0.078 & 0.105 & 0.075 \\
    \cmidrule(lr){1-10}

    \multirow{2}{*}{Elasticsearch}
    & F1 & 0.162 & 0.151 & 0.158 & 0.232 & 0.124 & 0.108 & 0.135 & 0.161 \\
    & RO & 0.114 & 0.078 & 0.041 & 0.022 & 0.065 & 0.015 & 0.062 & 0.055 \\
    \cmidrule(lr){1-10}

    \multirow{2}{*}{Loki}
    & F1 & 0.098 & 0.085 & 0.092 & 0.148 & 0.058 & 0.038 & 0.071 & 0.085 \\
    & RO & 0.108 & 0.088 & 0.058 & 0.042 & 0.092 & 0.038 & 0.078 & 0.068 \\
    \cmidrule(lr){1-10}

    \multirow{2}{*}{DIN-SQL}
    & F1 & 0.131 & 0.089 & 0.169 & 0.106 & 0.035 & 0.060 & 0.060 & 0.083 \\
    & RO & 0.050 & 0.016 & 0.066 & 0.136 & 0.027 & 0.039 & 0.053 & 0.022 \\
    \cmidrule(lr){1-10}

    \multirow{2}{*}{\textbf{LogNLQ}}
    & \textbf{F1} & \textbf{0.892} & \textbf{0.690} & \textbf{0.948} & \textbf{0.935} & \textbf{0.887} & \textbf{0.547} & \textbf{0.910} & \textbf{0.771} \\
    & \textbf{RO} & \textbf{0.854} & \textbf{0.504} & \textbf{0.774} & \textbf{0.425} & \textbf{0.776} & \textbf{0.643} & \textbf{0.705} & \textbf{0.331} \\
    \bottomrule
  \end{tabular}
\end{table*}

\runinhead{Log Retrieval}
LogNLQ leads all baselines by a wide margin in every setting: on basic queries, the best baseline reaches around 0.20 while LogNLQ exceeds 0.89; on scenario queries, baselines cluster below 0.25 while LogNLQ remains above 0.54 on all datasets. Among baselines, Elasticsearch occasionally achieves the highest F1 through broad inverted-index recall, but high recall does not translate into precision; text-based QA methods (LogQA, LogSay) score lowest throughout.

\runinhead{Metric Aggregation}
The RO rows of Table~\ref{tab:combined_performance_merged} report RO-Match for aggregation queries, which require jointly resolving event type, filter conditions, aggregation target, and grouping logic.
The performance gap between LogNLQ and baselines is even wider than in retrieval. On scenario aggregation, most baselines drop to near zero. LogQLLM is the strongest baseline overall, yet it still falls far below LogNLQ on both basic and scenario queries across all datasets.
The consistent advantage of LogNLQ on this more demanding task type confirms that it is capable of supporting complex grouped and multi-condition aggregations where all baselines effectively fail.

\runinhead{Error analysis}
Manual inspection of failed predictions reveals four recurring failure patterns.
(1)~\emph{Over-broad content matching}: raw-log Text-to-SQL and DIN-SQL predicates capture target keywords without preserving parameter roles, admitting structurally misleading matches (e.g., an IP appearing as a username satisfies a \texttt{LIKE} predicate).
(2)~\emph{Field and label hallucination}: Log-DSL Generation methods produce plausible but nonexistent field names over schema-light backends, causing empty or failed executions and preventing distinction between semantically similar parameters in the same log line.
(3)~\emph{Incorrect aggregation resolution}: without explicit parameter columns, methods cannot reliably identify grouping keys or aggregation targets under scenario-level complexity. LogSay lacks structured output capacity entirely, while DIN-SQL's schema-free generation fails to resolve the correct fields across varying template diversity.
(4)~\emph{LogNLQ retrieval misses}: LogNLQ's own failures concentrate in schema retrieval coverage, where the relevant template falls outside the top-$K$ candidates, or residual semantic ambiguity, where a user expression underspecifies which of multiple candidate columns is intended.

Patterns~(1)--(3) explain the baseline failures; LogNLQ's own errors (pattern~4) are qualitatively different, arising from retrieval coverage gaps rather than the absence of executable structure.

\subsection{RQ2: Efficiency and Overhead}

We next evaluate the overhead of LogNLQ in terms of storage footprint, offline build cost, and online latency. Our goal is not to compare against fully distributed production deployments, but to assess whether the LogNLQ design can be realized with practical overhead in a representative analytical implementation.

\begin{figure}[t]
  \centering
  \includegraphics[width=0.86\columnwidth]{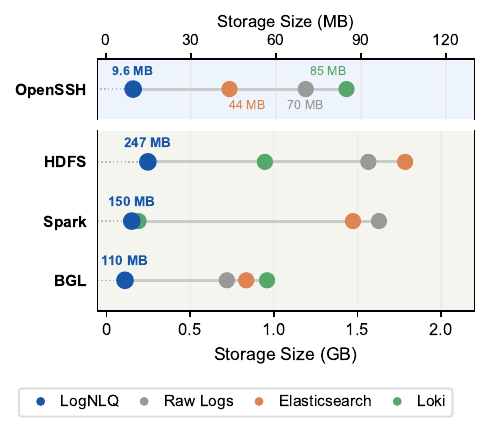}
  \caption{Storage Overhead Comparison.}
  \Description{LogNLQ consistently uses less storage than raw logs, Elasticsearch, and Loki.}
  \label{fig:storage_overhead}
\end{figure}

\begin{figure}[t]
  \centering
  \includegraphics[width=\columnwidth]{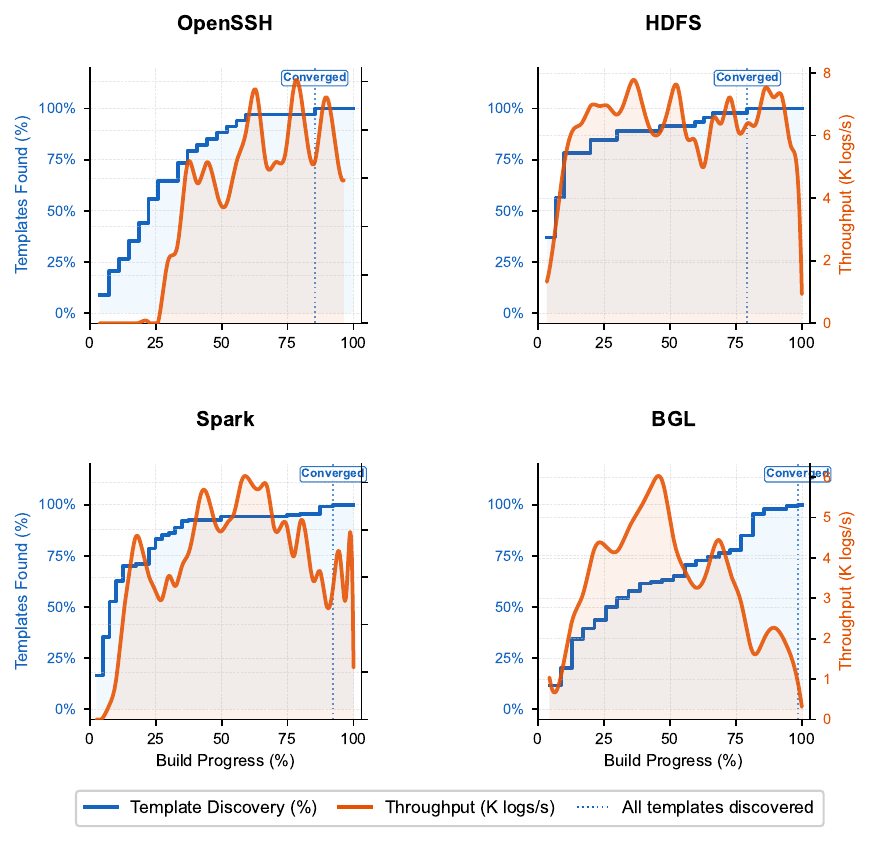}
  \caption{Offline Build Throughput.}
  \Description{Build throughput across four datasets, stabilizing after template discovery converges.}
  \label{fig:build-performance}
\end{figure}

\begin{figure}[t]
  \centering
  \includegraphics[width=0.92\linewidth]{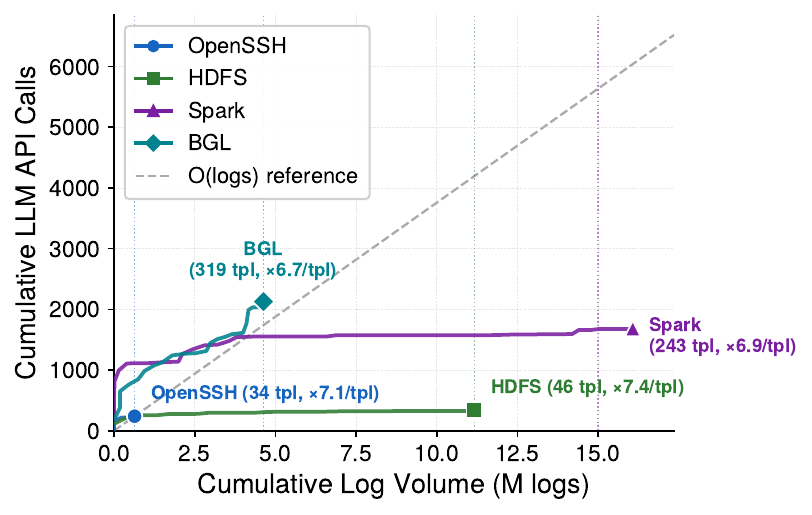}
  \caption{LLM API Call Scalability.}
  \Description{LLM API calls scale with unique template count, not with log volume.}
  \label{fig:llm-scalability}
\end{figure}

\runinhead{Storage footprint} 
Figure~\ref{fig:storage_overhead} compares the storage cost of LogNLQ with representative backend systems under two views: absolute size (log scale) and size relative to raw logs. The key finding is that LogNLQ consistently \emph{compresses} relative to raw logs, ranging from 33\% to 62\% of the original size, while Elasticsearch and Loki both \emph{expand} beyond the raw-log baseline (108\%--133\%). For example, on HDFS, LogNLQ requires 247\,MB, compared with around 1.5\,GB for Elasticsearch and 1.3\,GB for Loki. This result shows that explicit relational structuring, when paired with columnar compression, does not trade storage efficiency for richer structure; it actually reduces footprint relative to both raw-log and index-based backends.

\runinhead{Offline build throughput}
Figure~\ref{fig:build-performance} tracks per-interval throughput and cumulative template discovery across four datasets. Throughput starts near zero during the cold-start phase while new templates are registered, then rises sharply and stabilizes once the template vocabulary converges. The cold-start duration tracks template count: HDFS (46 templates) converges quickly while BGL (319 templates) takes longer. After convergence, the two largest datasets sustain throughputs of hundreds of thousands of lines per minute, completing on a single machine in approximately 30 minutes (HDFS, 11.2\,M lines) and 40 minutes (Spark, 16.1\,M lines).

Figure~\ref{fig:llm-scalability} shows that cumulative LLM API calls flatten well before the full corpus is ingested. The per-template call ratio is stable at approximately $7\times$ across all four datasets, meaning API cost can be estimated from the template vocabulary size alone. Once the vocabulary stabilizes, additional log lines incur zero extra cost.

\begin{figure}[t]
  \centering
  \includegraphics[width=\linewidth]{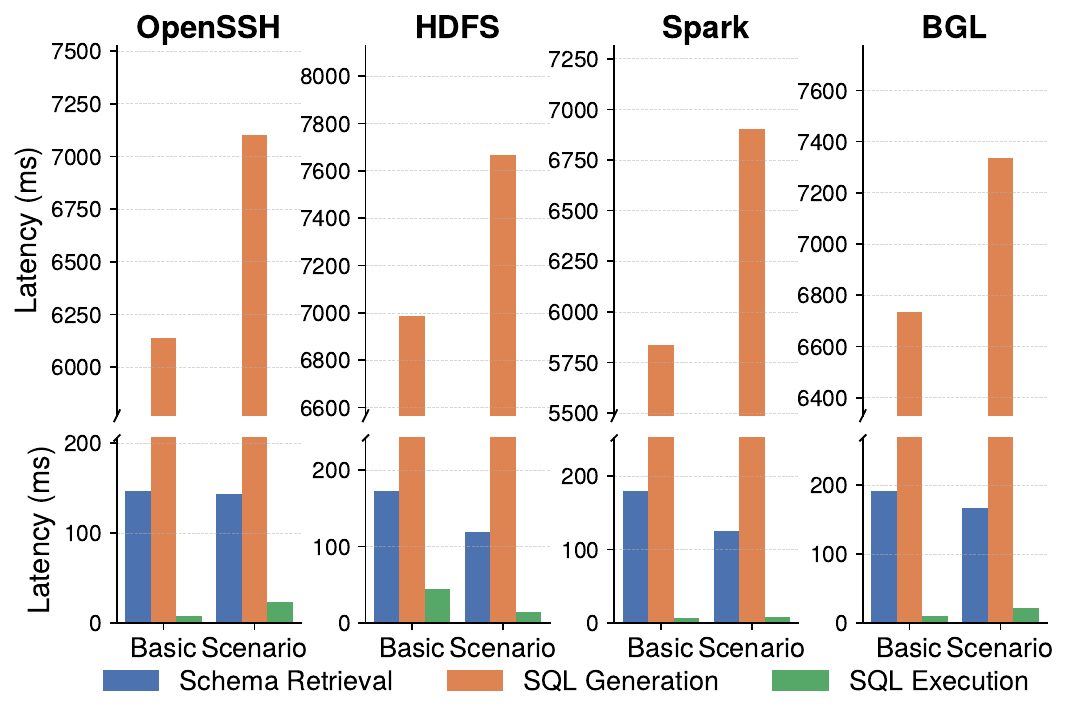}
  \caption{End-to-End Query Latency.}
  \Description{LLM inference dominates end-to-end latency, accounting for over 90\% of total query time across all datasets.}
  \label{fig:query_latency}
\end{figure}

\runinhead{Online latency}
Figure~\ref{fig:query_latency} breaks down end-to-end latency into three
components for basic and scenario queries across all four datasets.
Schema retrieval over FAISS takes 119--201\,ms and SQL execution over Parquet
takes 7--48\,ms; both remain negligible regardless of dataset or query type.
SQL generation by the LLM accounts for 5{,}692--7{,}669\,ms, or over 90\% of
the total. The overall latency range is narrow across four datasets spanning
very different log formats and scales, showing that the system's structural
overhead is stable and not sensitive to log diversity. The sole bottleneck is
LLM inference, which is external to our design and will improve as model
serving advances.

Taken together, these results show that LogNLQ can be realized with practical storage and latency overhead in a single-node analytical setting.

\subsection{RQ3: Ablation and Sensitivity}

\begin{figure}[htbp]
  \centering
  \includegraphics[width=\columnwidth]{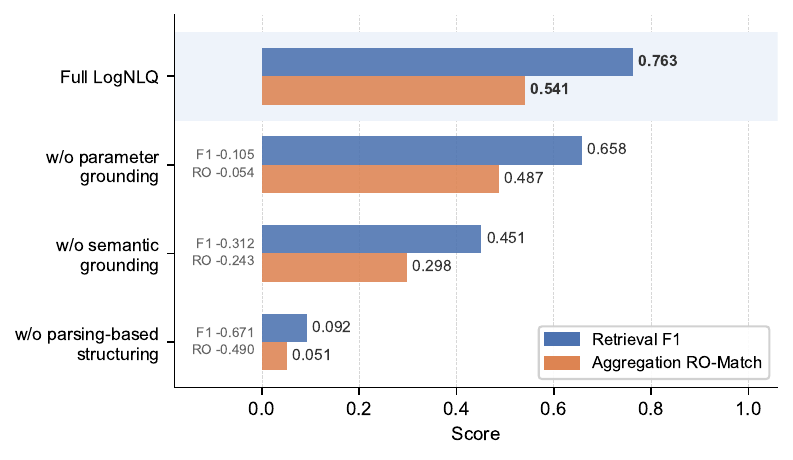}
  \caption{Component Analysis on Scenario Queries.}
  \Description{Performance drop per ablated component, averaged across four datasets on scenario queries.}
  \label{fig:ablation}
\end{figure}

\runinhead{Component analysis}
We examine the contribution of LogNLQ's main components by evaluating three ablation variants. Figure~\ref{fig:ablation} reports the average Retrieval F1 and Aggregation RO-Match across all datasets on scenario queries at $K{=}5$, with each variant's bars annotated by the performance drop ($\Delta$) relative to the full system:

\begin{itemize}[noitemsep,leftmargin=5.5mm]
    \item \textbf{w/o parameter grounding}: During online query prompt assembly, parameter semantic annotations (canonical name $n_i$, description $d_i$) and statistical profiles $\sigma_i$ are removed; only physical column names (\texttt{p\_1}, \texttt{p\_2}, \ldots) are retained.
    \item \textbf{w/o semantic grounding}: Templates are vectorized directly from their raw template strings $T_k$ for retrieval, and the prompt provides only the bare physical structure: event identifier $e_k$, the symbolic template string $T_k$, and original column names, with no semantic information of any kind.
    \item \textbf{w/o parsing-based structuring}: We equate this variant with the DIN-SQL baseline already reported in RQ1, which operates over a monolithic \texttt{Content} storage layout without any parser-induced structure or semantic index, and represents the strongest schema-free approach in our comparison.
\end{itemize}

The results reveal a clear hierarchical dependency among LogNLQ's three components. The full system achieves a Retrieval F1 of 0.763 and an Aggregation RO-Match of 0.621. The most striking finding is the near-complete collapse of aggregation when parsing-based structuring is removed. Aggregation RO-Match drops to 0.032 while Retrieval F1 falls to 0.092. This confirms that parser-induced relational structure is not merely a performance booster but an absolute prerequisite. Without explicit template partitions and parameter columns, the system cannot perform corpus-level analytical computation at all. Removing all semantic grounding (while retaining the parser-induced structure) causes the next-largest drop, with F1 falling by 0.312 and RO-Match by 0.243. This shows that executable structure alone is insufficient when the induced schema remains opaque. Positional columns such as \texttt{p\_1} and \texttt{p\_2} provide correct execution boundaries but cannot be reliably aligned to user intent without interpretable semantic annotations. Finally, removing only parameter-level grounding causes a more targeted degradation (F1 $-$0.105, RO-Match $-$0.054), with a proportionally larger impact on aggregation, consistent with the intuition that aggregation tasks require precise column-level alignment for correct grouping keys and filter targets.

\begin{figure}[t]
  \centering
  \includegraphics[width=\columnwidth]{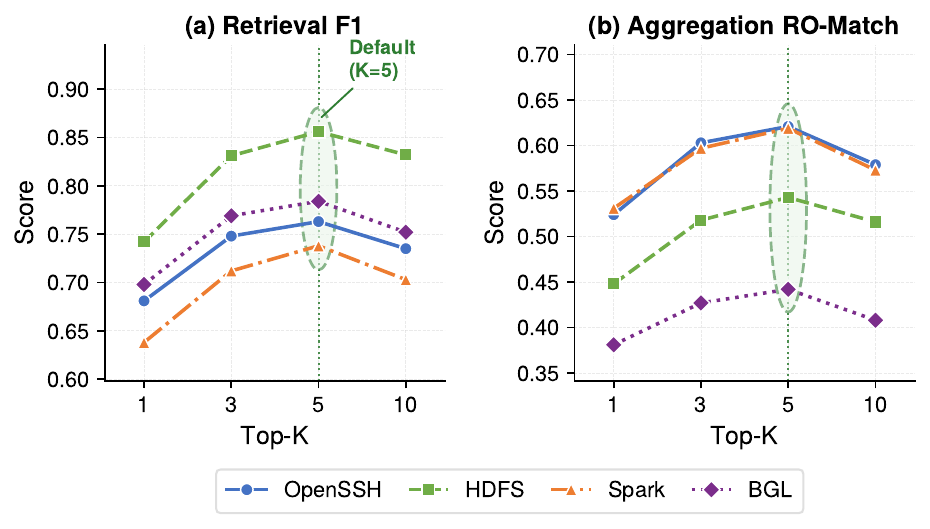}
  \caption{Top-$K$ Sensitivity.}
  \Description{Sensitivity to Top-$K$ retrieved schema candidates across four datasets on scenario queries.}
  \label{fig:sensitivity_topk}
\end{figure}

\runinhead{Top-$K$ sensitivity}
We evaluate sensitivity to the number of retrieved schema candidates across all four datasets. This parameter controls a trade-off in online schema selection: using too few candidates may omit the relevant induced template, while using too many may introduce distracting schema context into the generation prompt. As shown in Figure~\ref{fig:sensitivity_topk}, performance is best at moderate Top-$K$ values across all datasets, while both overly small and overly large settings lead to some degradation. Adding more candidates initially improves coverage but eventually dilutes the LLM's attention. At $K{=}5$, both Retrieval F1 and Aggregation RO-Match consistently peak across all four datasets, outperforming both $K{=}3$ and $K{=}10$.

\begin{figure}[t]
  \centering
  \includegraphics[width=\columnwidth]{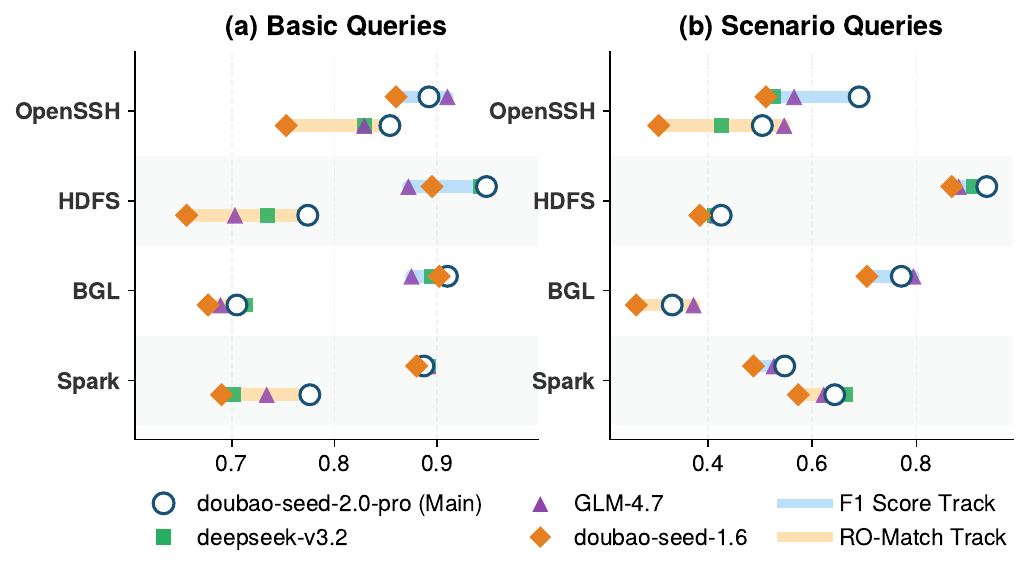}
  \caption{LLM Backbone Sensitivity.}
  \Description{Four LLM backbones compared; performance gap widens on scenario vs.\ basic queries.}
  \label{fig:backbone}
\end{figure}

\runinhead{Sensitivity to LLM Backbone}
Because LogNLQ relies on an LLM for schema-aware SQL generation, we also evaluate its sensitivity to the choice of foundation model. To ensure a fair and controlled comparison, all generative baselines in our main experiments share the \texttt{doubao-seed-2.0-pro} backbone. Here, we further examine how the choice of backbone affects performance by additionally instantiating LogNLQ with three alternative models: \texttt{deepseek-v3.2}, \texttt{GLM4.7}, and \texttt{doubao-seed-1.6}.

Figure~\ref{fig:backbone} illustrates the averaged performance of these models. \texttt{doubao-seed-2.0-pro} achieves the strongest results overall, with the most pronounced advantage on scenario queries and aggregation tasks that demand rigorous multi-step reasoning. For instance, replacing it with \texttt{doubao-seed-1.6} reduces the average Scenario F1 from 0.736 to 0.643 and Basic RO-Match from 0.777 to 0.694. As the dumbbell plots visually confirm, the performance gap between the reference model and alternative models is relatively narrow on basic queries, but widens substantially on complex diagnostic scenarios. Crucially, all tested backbones consistently and substantially outperform the baselines reported in RQ1, confirming that parser-induced structuring and semantic grounding remain the dominant enablers of executable log analytics, and that LogNLQ provides a robust, model-agnostic execution substrate whose accuracy scales natively alongside advances in foundational LLMs.

\subsection{Threats to Validity}

The primary threat to internal validity lies in the implementation of our proposed approach and the baselines. To mitigate this, we utilize author-provided implementations for the baselines and perform meticulous code reviews and testing of our own approach and experimental scripts.

Threats to external validity concern generalizability. We mitigate this by evaluating on four real-world log datasets from LogHub-2.0 (OpenSSH, HDFS, Spark, and BGL) that span varying system types and scales, and by covering both Basic and Scenario query workloads against baselines from three distinct paradigms: Text-based Log QA, Log-DSL Generation, and Text-to-SQL. Our evaluation also includes storage footprint, build throughput, and online latency metrics. Finally, while LogNLQ supports cross-template SQL generation, the quality of such queries depends on the LLM's multi-table reasoning capability and may degrade for highly complex join conditions; a systematic characterization is left for future work.

Threats to construct validity are mainly related to whether our evaluation metrics accurately reflect the effectiveness of the approach. To alleviate it, we adopt execution-verified metrics—F1 score for log retrieval and Result-Oriented Match (RO-Match) for metric aggregation—that evaluate the actual executed outputs rather than just the surface form of the generated queries. Furthermore, we conduct a manual error analysis to provide fine-grained insight into failure patterns across all methods.

\section{Related Work}
\label{sec:related_work}

\subsection{Automated Log Parsing and Schema Induction}
Automated log parsing~\cite{DBLP:journals/pacmse/HuangJCL25} is the foundational step for transforming semi-structured raw logs into structured data. Classical approaches, such as Drain~\cite{he2017drain}, rely on syntax-based heuristics to separate static templates from dynamic variables. Recently, the field has experienced a paradigm shift towards semantic and LLM-assisted parsing. For instance, DivLog~\cite{DBLP:conf/icse/XuYH0H24} and LibreLog~\cite{DBLP:conf/icse/MaKC25} leverage in-context learning to extract log templates without extensive rule engineering. To mitigate the latency bottlenecks of LLMs in production environments, frameworks like InferLog~\cite{DBLP:journals/corr/abs-2507-08523} and LILAC~\cite{jiang2024lilac} introduce adaptive caching mechanisms. Furthermore, recent large-scale evaluations~\cite{jiang2024large, DBLP:journals/pacmse/ZhangXLZLGZZ25} emphasize that semantic understanding is becoming the new standard for log parsing. 

While most of these recent advancements optimize parsing as a strict preprocessing step for downstream tasks—such as automated anomaly detection~\cite{DBLP:journals/pacmse/Xie0024, DBLP:conf/kbse/XiaoDHJWXGZHLH25} or multi-modal root cause analysis~\cite{DBLP:conf/kbse/WangZFMH24, DBLP:conf/issta/ShanH000Z24}—\alias\ takes a divergent approach. Rather than merely cleaning data for classifiers, our framework uniquely repurposes log parsing as an execution-oriented \emph{schema induction} mechanism, dynamically constructing relational tables to support analytical querying.

\subsection{Natural Language Interfaces for Log Analysis}
As modern systems generate logs at an unprecedented scale, researchers have explored various techniques to facilitate log management, ranging from advanced log compression~\cite{DBLP:conf/kbse/YuW0H24, shan2026logfold} to intelligent querying interfaces. Natural language querying (NLQ) over logs has gained traction to lower the barrier for developers and SREs. 

Existing approaches generally fall into two categories. The first is \emph{Text-based Log QA}, such as LogQA~\cite{huang2023logqa} and LogSay~\cite{qi2024logsay}. More recent explorations in this vein, including LogPrompt~\cite{DBLP:conf/icse/0001TMYZY24} and LogExpert~\cite{DBLP:conf/icse/WangC0SQWQL24}, investigate prompt engineering to enable zero-shot log analysis directly from text. While effective for localized troubleshooting, direct LLM-based QA often struggles with complex mathematical aggregations across millions of log lines and is susceptible to hallucination. The second category is \emph{Log-DSL Generation}, exemplified by LogQLLM~\cite{seshagiri2024chatting}, which translates user intents into domain-specific query languages. However, relying on custom DSLs often tightly couples the interface to specific backend architectures. 

\alias\ bridges the gap between these paradigms. By treating the parsed logs as a dynamic schema and employing dual-granularity semantic grounding, \alias\ empowers the LLM to act as a constrained SQL generator. This ensures that the analytical reasoning is offloaded to a deterministic, verifiable SQL execution engine, offering both the flexibility of natural language and the rigorous accuracy required for operational data analysis.

\section{Conclusion}
\label{sec:conclusion}

In this paper, we present \alias, a framework for natural-language log querying over parser-induced and semantically grounded schemas. By treating log parsing as an execution-oriented schema induction mechanism and complementing it with dual-granularity semantic grounding, \alias enables an LLM to act as a constrained SQL generator over induced relational structures rather than reasoning directly over raw log text. For rigorous evaluation, we also introduce LogNLQ-Bench, an execution-verified benchmark of 8,895 queries over four real-world datasets. Experiments validate the effectiveness of \alias. It consistently outperforms representative baselines across all datasets and query types, with especially clear gains on scenario-driven and analytically complex queries. In the future, we intend to extend this paradigm toward richer cross-system and multi-modal observability analysis involving logs, traces, and metrics, as well as more complex multi-hop diagnostic scenarios that require reasoning across heterogeneous event sources.

\section*{Data Availability}

The implementation of LogNLQ and LogNLQ-Bench are available on \url{https://doi.org/10.5281/zenodo.19251973}.

\bibliographystyle{ACM-Reference-Format}
\bibliography{references}

\end{document}